\documentstyle[aps,preprint,floats]{revtex}
\tightenlines
\begin{document}

\draft
\title{Weakly bound states in $2+\epsilon$ dimensions}
\author{S.M. Apenko\thanks{E-mail: apenko@td.lpi.ac.ru}}
\address{P.N. Lebedev Physical Institute, Moscow, 117924, Russia}
\date{\today }
\maketitle

\begin{abstract}
We study the critical behaviour near the threshold where a first
bound state appears at some value of coupling constant in an
attractive short-range potential in $2+\epsilon $ dimensions. We
obtain general expression for the binding energy near the threshold
and also demonstrate that the critical region is correctly
described by an effective separable potential. The critical
exponent of the radius of weakly bound state is shown to coincide
with the correlation length exponent for the spin model in the
large-$N$ limit. In two dimensions, where the binding energy is
exponentially small in coupling constant, we obtain a general
analytic expression for the prefactor.
\end{abstract}


\section{Introduction}

It is well known, that in three dimensions a bound state for a particle in a
short-range potential exists not at an arbitrary value of the coupling
constant $\lambda $, but only at $\lambda \geq \lambda _c$, where $\lambda
_c $ is a critical value which depends on the particular potential. This may
be viewed as the simplest example of a quantum phase transition, when e.g.
an excitation gap vanishes as some parameter of the Hamiltonian is varied
(see e.g. \cite{quantum} for a review). In a sense, such a behaviour is
similar to the second-order phase transition. Near the threshold the energy
of the bound state behaves like an `order parameter' $E\sim (\lambda
-\lambda _c)^\beta $ , where $\lambda $ plays the role of temperature and
$\beta $ is the critical exponent. Deeper investigation of this critical
behaviour is interesting in itself and may have some applications (see \cite
{Lassaut} and references therein).

The lower critical dimensionality for this transition is $d=2$, since in two
dimensions there always exists a bound state with energy exponentially small
in $\lambda $ \cite{Landau}, hence $\lambda _c=0$ in this case. For this
reason it seems natural to study the critical behaviour in $2+\epsilon $
dimensions, considering $\epsilon $ to be a small parameter, as it was done
for the phase transition in the nonlinear $O(N)$ sigma model \cite{Polyakov}
and also in the theory of Anderson localization (see e.g. \cite{Lee}). It
appears possible to develop an $\epsilon $-expansion both for the
wavefunction at the critical point and for the critical coupling $\lambda
_c(\epsilon )$ \cite{Apenko,Belov}. Even the first two terms of the
expansion of $\lambda _c(\epsilon )$ in powers of $\epsilon $ give rather
accurate estimate for $\lambda _c$ in three dimensions.

In this note we consider the onset of the first bound state in
$2+\epsilon $ dimensions in more detail. First, we demonstrate that
$\beta =2/\epsilon $ for $\epsilon \leq 2$ and $\beta =1$ above
four dimensions. The result at small $\epsilon $ is not unexpected,
since $\beta $ must go to infinity as $\epsilon \rightarrow 0$ to
reproduce the exponential dependence of $E$ on $\lambda $ in two
dimensions. This expressions for $\beta $ are consistent with the
results of Lassaut {\em et al} \cite{Lassaut}. Though they have
studied three dimensional case with non-zero orbital momentum $l$,
this is equivalent to the $s$-state problem in $2+\epsilon $
dimensions with $\epsilon =1+2l$ (see below).

However, the method used here is different from that of \cite{Lassaut}.
Starting from the integral representation of the Schr\"odinger equation we
first derive exact results for the binding energy at $\lambda \sim \lambda
_c $. Then we show also that the correct description of the critical region
is given by the {\em separable} approximation, with the true interaction
potential replaced by an effective nonlocal separable one, which depends on
the zero energy solution of the Schr\"odinger equation at $\lambda =\lambda
_c$.

When $\lambda $ $\rightarrow \lambda _c$ the radius of the bound
state diverges as $(\lambda -\lambda _c)^{-\nu }$ with the critical
exponent $\nu =\frac 12\beta $. Therefore the result obtained may
be represented in terms of $\nu $, namely $\nu =1/\epsilon $ at
$\epsilon \leq 2$ and $\nu =\frac 12$ at $\epsilon >2$. This
exponent coincides with the correlation length exponent in the
spherical model (equivalent to the $N$-component spin model at
$N\rightarrow \infty $)\cite{Ma} and with the localization length
exponent in the self-consistent theory of Anderson localization
\cite {Vollhardt}. It is not clear whether this coincidence imply
some nontrivial relation between these models, but still it seems
to be rather interesting.

The same critical exponents were obtained also from scaling
considerations by Hwa and Nattermann \cite{Hwa} and Kolomeisky and
Straley \cite{Kolomeisky}, who considered the problem of unbinding
a directed polymer from a columnar defect in the presence of
quenched disorder. Since such a polymer may be viewed as a
worldline of a quantum particle, in the clean case this problem is
essentially the same as the one discussed here.

Next, the approach used here also makes it possible to obtain an
asymptotic expression for the energy of the bound state in two
dimensions at $\lambda
\rightarrow 0$ along with the preexponential factor. This general analytic
expression for the prefactor seems, to the best of our knowledge, to be a
new one. We also calculate this prefactor for some simple potentials and
discuss its connection with the $\epsilon $-expansion for $\lambda _c$.

As a direct application of the general result for the binding energy in two
dimensions we also consider a case of a two-center potential. The attractive
force between the centers due to the bound state is shown to be of the
Coulomb type (compare with \cite{Lim}), and the universal prefactor is found
without solving the Schr\"odinger equation.

\section{Critical behaviour in $2+\epsilon$ dimensions.}

Consider the Schr\"odinger equation in $d$ dimensions

\begin{equation}
\label{S}-\Delta \Psi \left( {\bf r}\right) +\lambda V({\bf r})\Psi ({\bf r})=E\Psi ({\bf r}),
\end{equation}
( $\hbar =1$, $2m=1$), where $\lambda $ is the coupling constant
and $V({\bf r})$ is a short-range attractive potential. We assume
$V({\bf r)}$ to decrease faster then $1/r^2$ as $r\rightarrow
\infty $ and shall deal here with the weakly bound state of the
size $\sim (-E)^{-1/2}$ $\gg a$, where $a$ is the radius of the
potential.

The first bound state in the problem appears at some critical value of the
coupling constant $\lambda =\lambda _c$. Let us denote by $\psi _0$ the
wavefunction of this state at the threshold. Then $\psi _0$ obeys the
following zero energy equation

\begin{equation}
\label{zero}-\Delta \psi _0+\lambda _cV({\bf r})\psi _0=0
\end{equation}
It is convenient to normalize $\psi _0$ by the condition

\begin{equation}
\label{n}\int d{\bf r\,}\psi _0^2({\bf r})V({\bf r})=-1
\end{equation}
Note, that $\psi _0({\bf r})$ need not be square integrable and the
convergence of the normalization integral in (\ref{n}) is garantied by the
short-range potential $V({\bf r})$.

We shall now determine both the critical value $\lambda _c$ and the binding
energy at $\lambda \sim \lambda _c$. For this purpose first rewrite the
original Schr\"odinger equation in the integral form

\begin{equation}
\label{int}\Psi ({\bf r})=-\lambda \int d{\bf r}^{\prime }G_E({\bf r}-{\bf r}^{\prime })V({\bf r}^{\prime })\Psi ({\bf r}^{\prime })\,\,,
\end{equation}
where $G_E({\bf r})$ is the Green's function of the free particle. This
integral equation may be viewed also as an equation determining $\lambda (E)$
for the given negative energy $E$ of the bound state. Multiplying both sides
of (\ref{int}) by $V({\bf r})\Psi ({\bf r})$ and integrating over ${\bf r}$
we easily obtain

\begin{equation}
\label{lambda}\frac 1\lambda =-\frac{\int d{\bf r\,}d{\bf r}^{\prime }\,V({\bf r})\Psi ({\bf r})\,G_E({\bf r}-{\bf r}^{\prime })\,V({\bf r}^{\prime
})\Psi ({\bf r}^{\prime })}{\int d{\bf r\,}\Psi ^2({\bf r})\,V({\bf r})}
\end{equation}

If we put here $E=0$ then $\Psi \rightarrow \psi_0$ and using the
normalizing condition (\ref{n}), we have

\begin{equation}
\label{lam0}\frac 1{\lambda _c}=\int d{\bf r\,}d{\bf r}^{\prime }\,W({\bf r})\,G_0({\bf r}-{\bf r}^{\prime })W({\bf r}^{\prime })\,,
\end{equation}
where

\begin{equation}
W({\bf r})=V({\bf r})\psi _0({\bf r})\,.\
\end{equation}
At zero energy $G_0(r)$ is merely the Green function of the Laplace operator
and

\begin{equation}
\label{A5}G_0(r)={\frac 1{\epsilon \sigma _\epsilon }}\;{\frac 1{r^\epsilon }}
\end{equation}
(see e.g. \cite{Vladimirov}), where
\begin{equation}
\sigma _\epsilon ={\frac{2\pi ^{1+\epsilon /2}}{\Gamma (1+\epsilon /2)}}\
\end{equation}
is the area of the unit sphere in $2+\epsilon $ dimensions. Then the
critical value of the coupling constant may be represented as follows

\begin{equation}
\label{l}\lambda _c=\epsilon \;{\frac{2\pi ^{1+\epsilon /2}}{\Gamma
(1+\epsilon /2)}\;}\left[ \int d{\bf r\,}d{\bf r}^{\prime
}\frac{W({\bf r})\,W({\bf r}^{\prime })}{|{\bf r}-{\bf r}^{\prime
}|^{\epsilon}}\right]
^{-1}\;\ .
\end{equation}
This expresion explicitly demonstrate, that normally $\lambda _c$ tends to
zero as $\epsilon $ when we approach two dimensions.

Now, in the vicinity of the critical point we may write

\begin{equation}
\label{exp}\Psi =\psi _0+\delta \Psi \,,\qquad G_E=G_0+\delta G_E
\end{equation}
and assume all corrections to zero energy values to be small. If we
substitute (\ref{exp}) in (\ref{lambda}) and retain only terms of first
order in $\delta \Psi $ we finally obtain

\begin{equation}
\label{B}\frac 1\lambda =\frac 1{\lambda _c}+\int d{\bf r\,}d{\bf r}^{\prime
}\,W({\bf r})\delta G_E({\bf r}-{\bf r}^{\prime })W({\bf r}^{\prime })
\end{equation}
with $1/\lambda _c$ given by (\ref{lam0}). Note that the terms
containing $\delta \Psi $ cancel out. This cancellation is a
consequence of the zero energy equation
\begin{equation}
\psi _0({\bf r})=-\lambda _c\int d{\bf r}^{\prime }G_0({\bf r}-{\bf r}^{\prime })V({\bf r}^{\prime })\psi _0({\bf r}^{\prime })
\end{equation}
and is actually due to the right-hand side of (\ref{lambda}) being
a variational functional which is stable against small variations
of the true wavefunction \cite{Zubarev}. In the limit $E\rightarrow
0$ we can expand $\delta G_E({\bf r}-{\bf r}^{\prime })$ in
equation (\ref{B}) in powers of $E$. In $2+\epsilon $ dimensions
with $\epsilon <2$ we have at small negative $E$

\begin{equation}
\label{GE}\delta G_E\simeq -\frac{\Gamma (1-\epsilon /2)}{2\epsilon \pi }\left( -\frac E{4\pi }\right) ^{\epsilon /2}+{\cal O}(\,Er^{2-\epsilon }).
\end{equation}
(see Appendix A). Substituting this expression in (\ref{B}), we finally
obtain
\begin{equation}
\label{fin}E=-\left( A\,{\frac{\lambda -\lambda _c}\lambda }\right)
^{2/\epsilon }\ \qquad 0<\epsilon <2\,,
\end{equation}
where
\begin{equation}
\label{A}A=2^\epsilon \;{\frac{\Gamma (1+\epsilon /2)}{\Gamma (1-\epsilon /2)}\;\frac{\int d{\bf r}\,d{\bf r}^{\prime }\,W({\bf r})\,W({\bf r}^{\prime
})\,|{\bf r}-{\bf r}^{\prime }|^{-\epsilon }}{(\int d{\bf
r}\,W({\bf r}))^2}}\ \;.
\end{equation}

At $\epsilon >2$ , the leading term in the expansion (\ref{GE}) for $\delta
G_E(r)$ is $\sim E$ , and
\begin{equation}
\label{fin'}E\sim (\lambda -\lambda _c)\qquad \epsilon \geq 2\,.
\end{equation}
These equations formally solve the problem of the critical behavior
near the transition where the first bound state appears in
$d=2+\epsilon $ dimensions. In three dimensions equation
(\ref{fin}) leads to $E\sim (\lambda -\lambda _c)^2$ and for the
square well potential one can easily verify (using $\psi _0$ from
equation (\ref{well})of Appendix B) that equations (\ref{l}),
(\ref{A}) give the correct answer $\lambda _c=\pi ^2/4a^2$ and
$A=\pi ^2/8a$. Note, that if the Schr\"odinger equation is solved
in the critical point, i.e. $\lambda _c$ and $\psi _0$ are known,
one can also evaluate the prefactor $A$.

The results obtained should be compared with that of \cite{Lassaut}, where
the $l$-wave case was considered, because the radial $s$-wave Schr\"odinger
equation in $d=2+\epsilon $ dimensions is equivalent to the three
dimensional equation with non-zero orbital moment $l=(\epsilon -1)/2$ (see
Appendix A). Hence e.g. the dependence(\ref{fin}) is the same, as $E\sim
(\lambda -\lambda _c)^{2/(2l+1)}$ obtained in \cite{Lassaut} for $l<1/2$.

The particular form of equation (\ref{B}) suggests that the correct
description of the critical region near the threshold $\lambda \sim \lambda
_c$ can be obtained within the separable approximation. This approximation,
widely used in nuclear physics, involves replacing the original
potential $V({\bf r})$ with a non-local separable one, for which
the Schr\"odinger equation is exactly solvable. In our case one
should take

\begin{equation}
\label{sep}V_{sep}=-V\,|\psi _0\rangle \langle \psi _0|\,V\,,
\end{equation}
where $\psi _0$ is the zero energy solution normalized by the condition (\ref
{n}). If the ground state wave function $\Psi \approx \psi _0$ then $V_{sep}$
is in a sense close to $V$, since $(V-V_{sep})\psi _0=0$.

The Schr\"odinger equation for a particle in the potential (\ref{sep}) reads

\begin{equation}
\label{sr}-\Delta \Psi ({\bf r})-\lambda V({\bf r})\psi _0({\bf r})\int d{\bf r}^{\prime }
V({\bf r}^{\prime })\psi _0({\bf r}^{\prime })\Psi ({\bf r}^{\prime })=E\Psi ({\bf r})
\end{equation}
and has an obvious solution for the bound state, which up to a normalizing
constant is given by

\begin{equation}
\label{wf}\Psi ({\bf r})=-\,\lambda \int d{\bf r^{\prime }}\,G_E({\bf r-r^{\prime }})\,W({\bf r}^{\prime })\ ,
\end{equation}
The energy of the bound state is determined by substitution of (\ref{wf})
into equation (\ref{sr}) i.e. from the equation
\begin{equation}
\label{energy}1=\lambda \int d{\bf r}\,d{\bf r}^{\prime }\,W({\bf r})\,G_E({\bf r}-{\bf r}^{\prime })\,W({\bf r}^{\prime })\,.
\end{equation}
This is just the same equation as (\ref{B}), since $G_E=G_0+\delta
G_E$ and $1/\lambda _c$ is determined from (\ref{lam0}). Therefore
the separable approximation (\ref{sep}) results in exact
expressions (\ref{fin}), (\ref {fin'}) in the close vicinity of the
critical point. The validity of the separable approximation seems
to be due to the wavefunction (\ref{wf}) having correct asymptotic
behaviour at $r\gg a$. This is similar to the one dimensional case,
where the energy of the weakly bound state can be obtained
replacing the true $V(x)$ with a suitable $\delta $-function
potential, which also may be viewed as a separable one. In Appendix
B we show how one can naturally arrive at the separable potential
of the form (\ref{sep}).

As $\lambda \rightarrow \lambda _c$ the radius of the bound state
$\xi \sim (-E)^{-1/2}$ goes to infinity as $(\lambda -\lambda
_c)^{-\nu }$, where we have introduced a new critical exponent $\nu
$. Then from (\ref{fin}) and (\ref{fin'}) it follows that
\begin{equation}
\label{nu}\nu =\left\{
\begin{array}{cc}
\frac 1{d-2} & 2<d<4 \\
&  \\
\frac 12 & d\geq 4
\end{array}
\right.
\end{equation}

Critical exponent $\nu $ diverges as $d\rightarrow 2$ and `freezes'
above $d=4$ at the mean field value $\nu =\frac 12$. This is
precisely the correlation length exponent for the $N$-component
spin model at $N\rightarrow \infty $ \cite{Ma}. Another model where
equation (\ref{nu}) arises, is the self-consistent theory of
Anderson localization for a particle in a random potential. In this
case the localization length diverges as $(E_F-E_c)^{-v}$ if the
Fermi energy $E_F$ approaches the mobility edge $E_c$ and $\nu $ is
also given by equation (\ref{nu})\cite {Vollhardt}. It was even
argued that this result for $\nu $ is valid beyond the
self-consistent approximation and is in fact an exact one\cite
{Kunz,Suslov}.

This interesting coincidence arises from the fact that in all these models
resulting equation, determining the behaviour of the correlation length $\xi
$, has the form similar to (\ref{energy}) with $E\rightarrow \xi ^{-2}$. For
example, in the self-consistent theory of Anderson localization the
localization length $\xi $ is given by the equation
\begin{equation}
\label{loc}1=B\,E_F^{-2-\epsilon }\;\int_0^{q_0}\frac{dq\,q^{1+\epsilon }}
{q^2+\xi ^{-2}}\,,
\end{equation}
where $B$ is some constant and $q_0$ is a momentum cut-off \cite{Vollhardt}.
Comparing (\ref{loc}) with (\ref{energy}) we see that this is indeed the
equation for the binding energy $\xi ^{-2}$ in an effective short-range
separable potential with $\lambda \sim 1/E_F^{2+\epsilon }$, $a\sim 1/q_0$.
If the Fermi energy increases, $\lambda $ tends to zero and at $\lambda
=\lambda _c$ the bound state disappears. This critical point obviously
corresponds to the Anderson transition. Perhaps this is not a mere
coincidence and some direct mapping between these models might be
established.

To conclude this section we should like to mention that at small
$\epsilon $ the critical exponents derived here can be obtained
without actually solving the Schr\"odinger equation. This was done
e.g.  by  Hwa and Nattermann \cite {Hwa} and Kolomeisky and Straley
\cite{Kolomeisky}, who considered the problem of unbinding a
directed polimer (i.e. the worldline of a quantum particle) from a
columnar defect.  In this case simple scaling arguments immediately
lead to equation (\ref{nu}). In fact, one can take any quantity
(not necessarily the free energy as in \cite{Hwa,Kolomeisky}),
depending on some scale and look at the perturbation theory in
$\lambda $. Consider e.g. the Born series for the s-wave scattering
amplitude $f(k)$ (see e.g. \cite{Taylor})

\begin{equation}
f(k)=f_1(k)+f_2(k)+\cdots
\end{equation}
where at small $k$

\begin{equation}
\label{f}f_1\sim \lambda \tilde{V}(0)\,,\qquad f_2\sim \lambda ^2\int_0^\infty
dq\,q^{1+\epsilon }\,\frac{|\tilde{V}(q)|^2}{k^2-q^2+i0}\,.
\end{equation}
and $\tilde{V}(q)$ is the Fourier transform of the potential. For
the perturbation theory to be valid it is necessary that
$f_2/f_1\ll 1$. At small $\epsilon $ one has from (\ref{f})
$f_2\sim \lambda^2\tilde{V}^2(0)(1-(k/k_0)^\epsilon )/\epsilon $,
where $k_0\sim 1/a$, and hence the particle is essentially free on
a scale $k$ if

\begin{equation}
\label{scal}\frac{f_2}{f_1}\sim \frac{\lambda }
\lambda_c \left[ 1-\left(\frac k{k_0}\right) ^\epsilon \right] \ll 1
\end{equation}
(compare with \cite{Hwa}), where $\lambda_c \sim \epsilon
\tilde{V}(0)$. If $\lambda >\lambda_c$ then (\ref{scal}) is not fulfilled at $k=0$ and the
particle is bound. But on a length scale

\begin{equation}
\label{ksi}k\gg \xi ^{-1}\sim k_0\left( \frac{\lambda -\lambda _c}\lambda
\right) ^{1/\epsilon }
\end{equation}
one can neglect the potential. Hence $\xi $ from (\ref{ksi}) may be
viewed as the radius of the bound state. The critical exponent
obtained in this way is the same as (\ref{nu}).

\section{Weakly bound states in two dimensions.}

In three dimensions neither $\lambda _c$ nor $A$ are known exactly for an
arbitrary potential, since we can solve the zero-energy problem only in some
special cases. But the situation is different in two dimensions. In this
case arbitrarily weak short-range attractive potential binds a particle, so
that $\lambda _c=0$. Then the solution to zero-energy equation (\ref{zero})
is obviously $\psi _0(r)={\rm const}$, which, according to the normalization
condition (\ref{n}), results in $\psi _0=|\int d{\bf r}\,V({\bf r})|^{-1/2}$
and
\begin{equation}
W({\bf r})={\frac{V({\bf r})}{|\int d{\bf r}\,V({\bf r})|^{1/2}}}
\end{equation}

Next, in the limit $\epsilon \rightarrow 0$ we have $\lambda _c\sim \epsilon
$ and $A\rightarrow 1$. Therefore in this limit the right-hand side of
equation (\ref{fin}) for the binding energy turns to an exponential
function. Then for $\kappa =\sqrt{-E}$ we have
\begin{equation}
\label{E}\kappa =C\exp \left( -{\frac {2\pi}{\lambda |\int d{\bf r}\,V({\bf r})|}}
\right) \ ,
\end{equation}
where the prefactor $C$ is determined from the expansion of $A$ in
powers of $\epsilon $
\begin{equation}
\label{ac}A\simeq 1+\epsilon \ln C+\cdots \,.
\end{equation}
Expanding $A$ from (\ref{A}) in $\epsilon $ we obtain
\begin{equation}
\label{C}C=\exp \left( \ln 2-\gamma -{\frac{\int d{\bf r}d{\bf r}^{\prime
}\,V({\bf r})\,V({\bf r}^{\prime })\,\ln |{\bf r}-{\bf r}^{\prime
}|}{(\int d{\bf r}\,V({\bf r}))^2}}\right) \ ,
\end{equation}
where $\gamma =0.577\ldots $ is the Euler's constant. Thus, in two
dimensions we have a general explicit expression for the energy of
the weakly bound state. In contrast to the one dimensional case,
where at $a\rightarrow 0$ we may approximate any short range
potential by the $\delta $-function and the binding energy depends
only on one potential dependent integral $\int dxV(x)$, here there
are two different integrals, one of which being nonlocal. In three
dimensions no such general closed form for the energy is available
even near the threshold $\lambda =\lambda _c$.

The double integral in (\ref{E}) resembles the one encountered earlier in
the $\epsilon $-expansion for $\lambda _c(\epsilon )$. For the spherically
symmetric potential at small $\epsilon $ one has \cite{Apenko}
\begin{equation}
\label{lam}\lambda _c(\epsilon )\simeq \lambda _1\,\epsilon +\lambda
_2\,\epsilon ^2+\cdots \ ,
\end{equation}
where
\begin{equation}
\label{l'}\lambda _1=-{\frac 1{\int_0^\infty drr\,V(r)}}\ ,
\end{equation}
\begin{equation}
\label{l''}\lambda _2=-\,{\frac 12}\;{\frac{\int_0^\infty drr\int_0^\infty
dr^{\prime }r^{\prime }\,V(r)\,V(r^{\prime })\,\ln
{\frac{r_{>}}{r_{<}}}}{[\int_0^\infty drr\,V(r)]^3}}
\end{equation}
( $r_{>}$ ($r_{<}$) is the greater (lesser) of $r$, $r^{\prime }$). Then,
after some straightforward calculations, we obtain another expression for $C$
\begin{equation}
\label{C'}C={\frac 1{\bar a}}\;\exp \left( \ln \,2-\gamma -{\frac{\lambda _2}
{\lambda _1}}\right) \ ,
\end{equation}
where $\bar a$ is the mean range of the potential, defined by
\begin{equation}
\label{a}\ln \bar a={\frac{\int_0^\infty drr\,\ln {r}\,V(r)}{\int_0^\infty
drr\,V(r)}}\ .
\end{equation}

We evaluate $\lambda _1$, $\lambda _2$, $\bar a$ and $C$ for several widely
used potentials and the results are displayed in Table 1. Note that the
values of the prefactor $C$ are surprisingly simple.

As mentioned earlier \cite{Apenko}, equation (\ref{lam}) may be extrapolated
to $\epsilon =1$ to give a rather good estimate for the critical coupling in
three dimensions. For the Yukawa potential this results e.g. in $\lambda
_ca^2\simeq 1+\ln {2}\simeq 1.693\ldots $, which is close to the exact
result $\lambda _ca^2=1.6798\ldots $.

It is also possible to use $\epsilon $-expansion in the same manner to
evaluate the prefactor $A$ in three dimensions. From (\ref{ac}) it follows
that $A^{1/\epsilon }\simeq C$ at small $\epsilon $. Extrapolating this
result to $\epsilon =1$ we see that actually $C$ may be treated as a first
approximation for $A$ in three dimensions. For the square well potential
this approximation gives $Aa\simeq 2e^{-\gamma +1/4}\simeq $ $1.442\ldots $
the exact value being equal to $Aa=\pi ^2/8=1.234\ldots $. While
qualitatively correct, this first approximation is not very accurate.

\begin{table}[tbfp] \centering
\caption[table]{ Values for $\lambda_1$ and
$\lambda_2$ from (\ref{lam}),  $\bar{a}$ from (\ref{a}) and the
prefactor $C$ from (\ref {C}) for several different potentials
$V(r)$.}
\begin{tabular}{ccccc}
\hline
$-V(r)$ & $\lambda _1\,a^2$ & $\lambda _2\,a^2$ & $\bar a/a$ & $C\,a$ \\
\hline
$\exp (-r/a)$ & 1 & $\ln 2-\frac 14$ & $e^{1-\gamma }$ & $e^{-3/4}$ \\
$\theta (a-r)$ & 2 & $\frac 12$ & $e^{-1/2}$ & $2\,e^{-\gamma +1/4}$ \\
$(a/r)\exp (-r/a)$ & 1 & $\ln 2$ & $e^{-\gamma }$ & 1 \\
$\exp (-r^2/a^2)$ & 2 & $\ln 2$ & $e^{-\gamma /2}$ & $\sqrt{2}e^{-\gamma /2}$
\\ \hline
\end{tabular}
\end{table}

Equation (\ref{E}) is valid for spherically nonsymmetric potentials as well,
provided the range of the wavefunction $\kappa ^{-1}$ is much larger, than
the radius $a$ of a potential. Hence equation (\ref{E}) may be used in the
problem of several attractive centers.

In the case of two identical centers separated by a distance $R$ we may
write
\begin{equation}
V({\bf r)=}v{\bf (r)+}v{\bf (r+R)\,,}
\end{equation}
and upon substituting this potential in (\ref{C}) we obtain that $C\sim \exp
(-\frac 12\ln R)$ at $R\gg a$ for arbitrary short-range $v({\bf r)}$. Then
for the energy $E_2$ of a weakly bound state at $a\ll R\ll \kappa ^{-1}$
equations (\ref{E}) and (\ref{C}) yield
\begin{equation}
\label{two}E_2=-\,2\,e^{-\gamma }\;\frac{{\kappa }_0}R\ ,
\end{equation}
where $\kappa _0$ is the square root of the binding energy on one center
(given by (\ref{E}) and (\ref{C}) with the replacement $V({\bf r)\rightarrow
\,}v{\bf (r)}$). This is obviously the energy of the symmetric state. Note,
that the energy $E_2$ depends on the details of the interaction only through
$\kappa _0$.

It is interesting that the effective long range force between two centers,
resulting from the bound state is of the Coulomb type. In three dimensions
the corresponding energy is known to behave as $1/R^2$ \cite{Fonseca}. This
is related to the collapse of three particle system with zero-range
interaction, known as Thomas effect \cite{Thomas}. Less singular behaviour
of the energy (\ref{two}) at small $R$ is related to the absence of the
Thomas effect in two dimensions \cite{Adhikari}. For separable potential of
a particular type the dependence $\kappa _0/R$ in two dimensions was derived
in \cite{Lim}. We see now that this formula is quite universal.

\section{Acknowledgments}

The author is grateful to V V Losyakov, D A Kirzhnitz and A V
Klyuchnik for valuable discussions, to P I Arseev for critical
reading of the manuscript and to the unknown referee for pointing
out the references \cite {Hwa,Kolomeisky}. This work was supported
by the RBRF Grant 96-15-96616.

\appendix

\section*{A}

For the spherically symmetric potential the radial part of the $2+\epsilon $
dimensional Schr\"odinger equation reads
\begin{equation}
\label{Se}\left( -{\frac 1{r^{\epsilon +1}}}{\frac d{d\,r}}r^{\epsilon +1}{\frac d{d\,r}}+\lambda V(r)\right) \Psi (r)=E\Psi (r)
\end{equation}
The substitution $\Psi =\phi r^{-(1+\epsilon )/2}$ puts equation (\ref{Se})
in the form
\begin{equation}
\label{A2}\left( {\frac{d\,^2}{d\,r^2}}-{\frac{(\epsilon ^2-1)}{4r^2}}
-\lambda V(r)+E\right) \phi =0
\end{equation}
This is obviously the radial equation for the wave function with non-zero
angular momentum $l$ in three dimensions with $l(l+1)=(\epsilon ^2-1)/4$,
i.e. $l=(\epsilon -1)/2$.

Next we proceed to the evaluation of the free particle Green's
function, $G_E(r)$, in $2+\epsilon $ dimensions. This function
certainly can be found elsewhere in the literature but, for the
sake of completeness we give here its short derivation. The Green's
function satisfies the equation
\begin{equation}
\label{A3}(-\Delta -E)G_E({\bf r})=\delta ({\bf r})
\end{equation}
At ${\bf r}\neq 0$ we have for $f=r^{\epsilon /2}G_E(r)$
\begin{equation}
\label{A4}\left( {\frac{d\,^2}{dr^2}}+{\frac 1r}{\frac d{dr}}-
{\frac{\epsilon ^2}{4r^2}}+E\right) f=0
\end{equation}
Then, at $E<0$, $f(r)$ (which goes to zero as $r\rightarrow \infty $) is
proportional to the modified Bessel function $K_{\epsilon /2}(\kappa r)$,
where $\kappa =\sqrt{-E}$, hence
\begin{equation}
\label{gr}G_E(r)\sim r^{-\epsilon /2}\;K_{\epsilon /2}(\kappa r)\ .
\end{equation}
At $E=0$ the Green's function reduces to the fundamental solution
of the Laplace equation, which in $2+\epsilon $ dimensions is given
by equation (\ref{A5}). At $\kappa r\rightarrow 0$ and $\epsilon
<2$ we have
\begin{eqnarray}
K_{\epsilon /2}(\kappa r) &=&{\frac \pi 2\;\frac{I_{-\epsilon /2}(\kappa
r)-I_{\epsilon /2}(\kappa r)}{\sin (\epsilon \pi /2)}}=\;  \label{assy} \\
\ &=&{\frac \pi {2\sin (\epsilon \pi /2)}}\;\left[ {\frac{(\kappa
r/2)^{-\epsilon /2}}{\Gamma (1-\epsilon /2)}}-{\frac{(\kappa r/2)^{\epsilon
/2}}{\Gamma (1+\epsilon /2)}}+{\cal O}((\kappa r)^{2-\epsilon /2})\right] \,,
\nonumber
\end{eqnarray}
Using this asymptotics, identity $\Gamma (z)\Gamma (1-z)=\pi /\sin (\pi z)$
and comparing (\ref{gr}) with (\ref{A5}) we can fix unknown constant in (\ref
{gr})
\begin{equation}
\label{Afin}G_E(r)=\frac 1{(2\pi )^{1+\epsilon /2}}\,
(\frac \kappa r)^{\epsilon /2}\,K_{\epsilon /2}(\kappa r)\,.
\end{equation}
Small $\kappa $ behaviour of (\ref{Afin}) gives rise to the expansion (\ref
{GE}) for $\delta G_E=G_E-G_{0\text{ }}$in the main text.

\section*{B}

To derive the separable potential (\ref{sep}) one can start from the
following separable decomposition of a local potential $V({\bf r)}$
\begin{equation}
\label{separ}V({\bf r})=\sum_n\;\sigma _n\,V|\psi _n\rangle \langle \psi
_n|V\,,
\end{equation}
where the set of functions $|\psi _n\rangle $ is determined from the
eigenvalue equation
\begin{equation}
\label{eig}(\Delta +E)^{-1}V\psi _n=\eta _n(E)\psi _n\,,
\end{equation}
and $\sigma _n=\pm 1$ depending on the sign of $\langle \psi _n|V|\psi
_n\rangle $ (see e.g. \cite{Brown}). For negative $E$ this equation is in
fact the Schr\"odinger equation for the bound states, where $\psi _n$ is the
wave function of a bound state with energy $E$ in the potential $1/\eta
_n(E)\,V({\bf r})$, i.e.
\begin{equation}
\label{psi}(-\Delta +1/\eta _n\,V({\bf r}))\psi _n=E\psi _n\,.
\end{equation}
From this equation one can derive a normalizing condition, which in this
case is known to be $\langle \psi _n|V({\bf r})|\psi _m\rangle =\sigma
_n\delta _{nm}$ \cite{Brown}.

Since we are interested in the weakly bound state with $E\simeq 0$, we may
now take the limit $E\rightarrow 0$ in equation (\ref{psi}), i.e. we may
define the set $\psi _n$ with respect to the zero energy. In this case $\psi
_n$'s are the threshold wavefunctions and the corresponding values of
$1/\eta _n(0)$ are critical values of coupling constant. The
wavefunction $\psi _0$ of the first ground state at $E\rightarrow
0$ obeys the zero energy equation (\ref{zero}), where $\lambda
_c=1/\eta _0\left( 0\right) $ and $\eta _0$ is the largest
eigenvalue in (\ref{eig}). We do not know $\psi _n$'s exactly for
an arbitrary potential, but e.g. for the attractive square well of
radius $a$ in three dimensions one can easily obtain for the
$s$-states

\begin{equation}
\label{well}
\begin{array}{c}
\psi _n\sim \left\{
\begin{array}{cc}
\frac 1r\sin \sqrt{\lambda _n}r & \;r<a \\
(-1)^{n}\frac 1r & \;r>a
\end{array}
\right. \\
\\
\lambda _n=1/\eta _n=(\frac 12\pi +\pi n)^2\frac 1{a^2}\,.
\end{array}
\end{equation}

In the critical region $\lambda \sim \lambda _c$ the wavefunction
of the first bound state is very close to $\psi _0$, so it seems
natural to retain only the term with $\psi _0$ in the expansion
(\ref{separ}) as a zero approximation. This approach is similar to
the pole approximation in the scattering theory, valid for the
resonance scattering when there exist a weakly bound state.

Assume next that $\langle \psi _0|V|\psi _0\rangle \leq 0$ so that $\sigma
_0=-1$. Then, in the vicinity of the critical point, where the transition
from zero to one bound state occurs, we arrive at the separable
potential (\ref{sep})
\begin{equation}
V\simeq V_{sep}=-\;V|\psi _0\rangle \,\langle \psi _0|V\,\,,
\end{equation}
where $\psi _0$ is normalized by the condition (\ref{n}) .

\end{document}